\documentclass[aps,prl,twocolumn,showpacs,superscriptaddress,groupedaddress]{revtex4}  

\usepackage[latin1]{inputenc}
\usepackage{epsfig}
\usepackage{enumerate}
\usepackage{subfigure}
\usepackage{epstopdf}
\usepackage{color}
\usepackage[export]{adjustbox}

\begin{document}

\title{Nonsynchronous updating in the multiverse of cellular automata}
\author{Sandro M. Reia}
\affiliation{Faculdade de Filosofia, Ci\^encias e Letras de Ribeir\~ao Preto, Universidade de S\~ao Paulo, Ribeir\~ao Preto, S\~ao Paulo, Brazil}
\author{Osame Kinouchi}
\affiliation{Faculdade de Filosofia, Ci\^encias e Letras de Ribeir\~ao Preto, Universidade de S\~ao Paulo, Ribeir\~ao Preto, S\~ao Paulo, Brazil}

\begin{abstract}

	In this paper we study updating effects on cellular automata rule space.
	We consider a subset of 6144 order-3 automata from the space of 262144 bi-dimensional outer-totalistic rules.
	We compare synchronous to asynchronous and sequential updatings.
	Focusing on two automata, we discuss how update changes destroy typical structures of these rules.
	Besides, we show that the first-order phase-transition in the multiverse of synchronous cellular automata, revealed with the use of a recently introduced control parameter, seems to be robust not only to changes in update schema but also to different initial densities.
	

\end{abstract}


\maketitle

\section{Introduction}

	
	The cellular automata (CA) was created in the 1940s by Stanislaw Ulam and John von Neumann to study self-reproducing biological systems.
	Since then, deterministic CA gave origin to a specific field of study itself.
	In particular, Wolfram~\cite{wolfram01,ref16} has shown the whole variety of behaviors and patterns exhibited by automata rules.  
	He argued that any CA rule could belong to one of the following behavioral classes: fixed point (Class I), periodic (Class II), chaotic (Class III) and complex (Class IV).
	Besides, Langton~\cite{selflangton} claimed the existence of a control parameter ($\lambda$) that would classify CA rules in general and reveal a phase transition in CA rule space if plotted against a suitable but yet ``unknown'' order parameter.

	
	A recent attempt to determine the type of phase transition CA rules are related to was made by Reia and Kinouchi~\cite{Reia_Kinouchi2014}.
	We have proposed a control parameter $\sigma$ which classifies the $2$ dimensional ($2D$) automata rules in two regimes: a dead absorbing state (or vacuum) and a regime of high density of alive sites.
	The transition between both regimes was found to be of first-order with a metastable region, and the most interesting rules.	
	We like to think, along other authors, that the rule space, with its diverse dynamical laws, represent a toy model for the multiverse~\cite{SolerGil2013}.
	In this sense, we found that complex ``universes'' like Game of Life~\cite{ref14} are special because they lie in the metastable region of the transition, which is a better description than the edge of chaos concept for CA~\cite{Langton1990}.

	
	Cellular automata are not only used to study critical properties of deterministic abstract dynamical systems, but also to study natural and social phenomena.
	In general, CA can be applied where there are interest on dynamical evolution, global pattern formation and statistical properties of multiple interacting agents.
	This agent-base modelling spread over a wide range of science domains and is very important to computational simulations of complex systems.
	In these cases, however, the automaton is usually probabilistic.
	We find, for instance, applications from ferromagnetism, like Ising~\cite{Ising_Model} and Blume-Capel\cite{Blume_Capel1974} models, to social interaction modelling, like majority vote model~\cite{majority_vote_model} and Axelrod model~\cite{Axelrod1997}.
	Examples of further applications of probabilistic CA can still be found in~\cite{Dickman1999,Castellano2009}.
		
	
	The equivalence between computer simulations and real problems was addressed by Huberman and Glance~\cite{Huberman1995}.
	These authors emphasize that one must be concerned about the effects of the update procedure on the global results of the simulation.
	According the authors, this concern is necessary to obtain a reliable insight into the real problem.
	Specifically, Huberman and Glance studied the Prisioner's Dilemma in a square lattice with synchronous and asynchronous updating and showed the results strongly differ.

	
	In this perspective, this paper presents a study of the effects of updating schemas and different initial densities on the CA phase transition reported in~\cite{Reia_Kinouchi2014}.
	In particular, we study how asynchronous and sequential updating affect the $2D$ CA rule space.
	Regarding the results reported in~\cite{Fates2010}, in which authors investigate how Game of Life steady states are related to its initial density and asynchronity, we also study the stability of $2D$ patterns which have similar densities to the fixed point densities predicted by mean field analysis.
	Our results suggest that although nonsynchronous updating destroys characteristic structures found in synchronous updating, which is in agreement with~\cite{Huberman1995,blok1},and that although the initial density can act in favor of dead or alive domains, the first-order phase-transition found in~\cite{Reia_Kinouchi2014} seems to remain unchanged.
	

\section{Rules and Updating Schemas}


	We will work in the rule space of binary outer-totalistic $2D$ CA. 
	The most known example of such rules is the Game of Life, where each site is either dead ($0$) or alive ($1$).
	From any initial configuration, the state of a given site is updated according its actual state $s$ and the number $h$ of alive neighbors in its Moore neighborhood.
	Precisely, a dead site becomes alive (``born $B$'') only if it has three alive neighbors, and an alive cell remains alive (``survive $S$'') only if it has two or three alive neighbors - all other situations lead to a dead site at next time step.
	An usual code for this rule is $B3S23$.

\begin{figure}[b!]
	\centering
		\subfigure[]{\includegraphics[width=0.15\textwidth,frame]{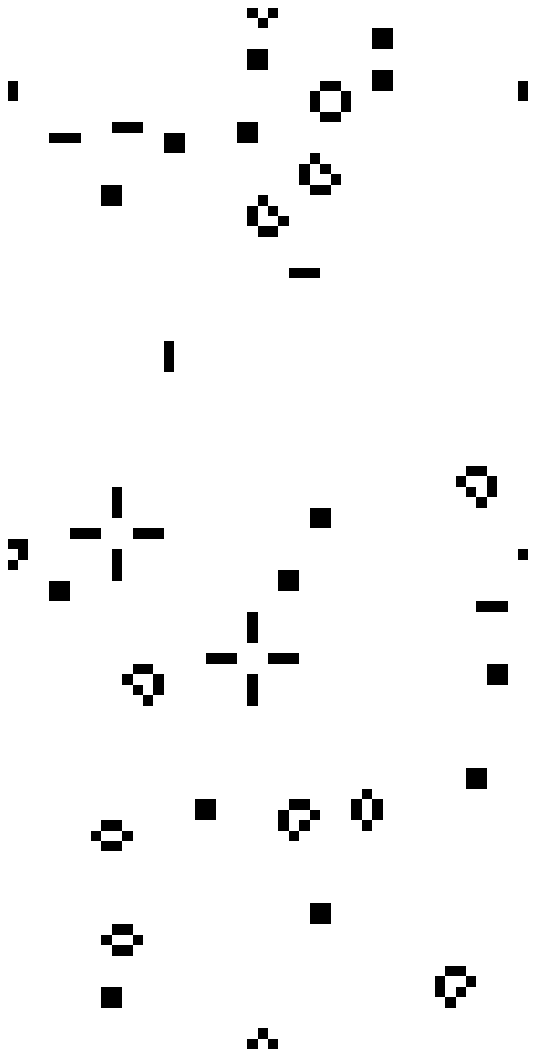}\label{LIFE_PARALELO}}
		\subfigure[]{\includegraphics[width=0.15\textwidth,frame]{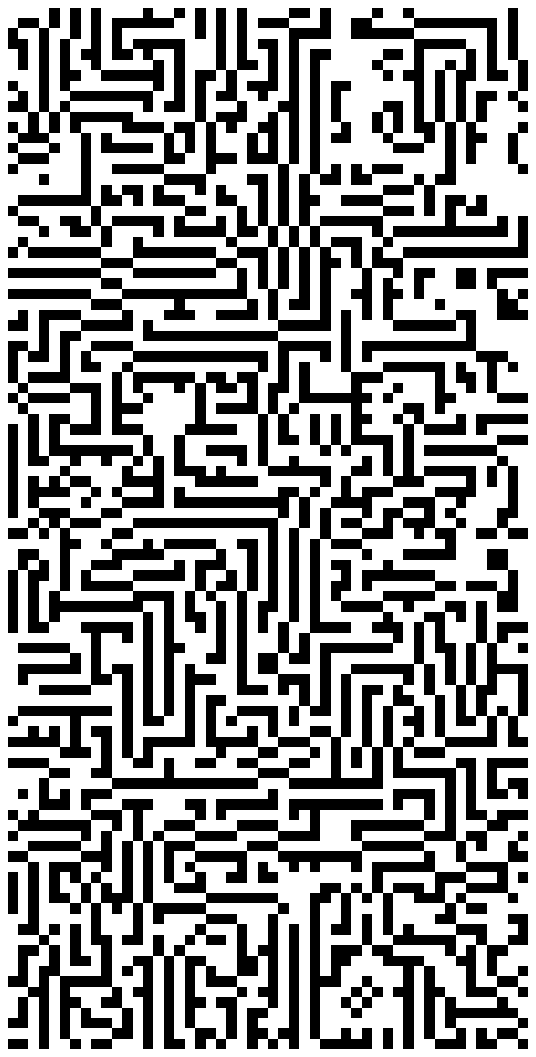}\label{LIFE_ASSINCRONO}}
		\subfigure[]{\includegraphics[width=0.15\textwidth,frame]{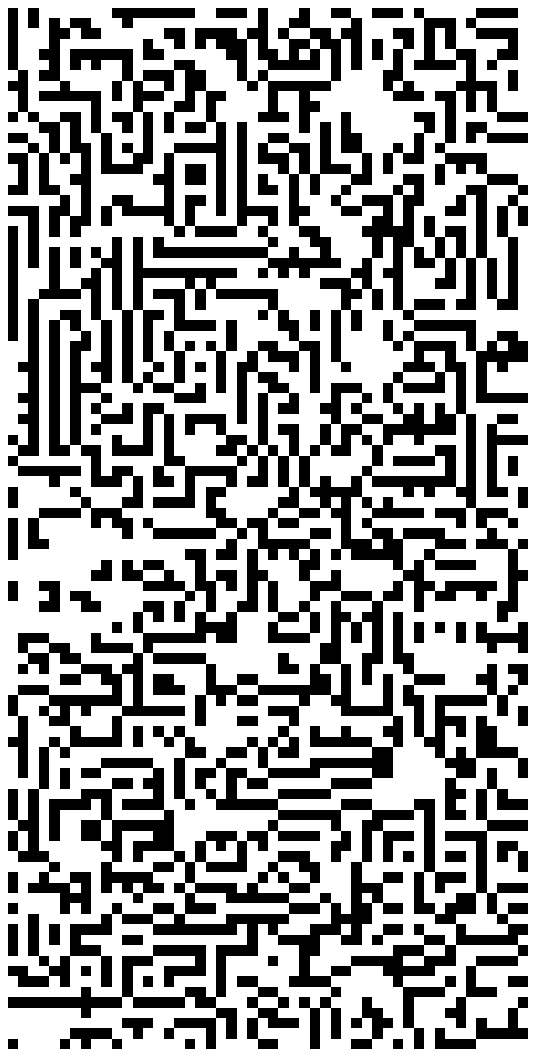}\label{LIFE_SEQUENCIAL}}
		\caption{Snapshot at steady state of Life with synchronous~\ref{LIFE_PARALELO}, asynchronous~\ref{LIFE_ASSINCRONO} and sequential~\ref{LIFE_SEQUENCIAL} updates in a $2D$ lattice with $L_m \times L_n=100 \times 50$. The system evolved to steady state from a randomly initial condition with $\rho(0)=0.5$. }
	\label{UPDATES_PATTERN}
\end{figure}

	
	Our results are obtained in a $2D$ lattice with periodic boundary conditions.
	All results reported here are from the steady steady of the system, at $t=5000$.
	The influence of update schemas can be viewed in Fig.~\ref{UPDATES_PATTERN}.
	In Fig.~\ref{LIFE_PARALELO} we have the usual synchronous update which leads to the well known structures of Game of Life~\cite{Bagnoli1991}.

	The implication of an asynchronous update for Life is seen in Fig.~\ref{LIFE_ASSINCRONO}.
	The asynchronous update is realized by choosing a site randomly and updating it according to the Life's rule.
	In this case, a time step is composed of $N=L_m \times L_n$ interactions and we see that the pattern is a labyrinth phase: the Life's characteristic configurations found in synchronous update are destroyed and a different pattern is established.
	We notice the number of alive sites is higher than seen in Fig.~\ref{LIFE_PARALELO}.
	This suggests that asynchronous updating acts in favor of alive sites by allowing the formation of a stationary labyrinth phase with $\rho_{\infty}=0.436$.
	A more detailed study about properties of Life under asynchronous updating can be found in~\cite{Fates2010,blok1}.

	
	Fig.~\ref{LIFE_SEQUENCIAL} shows a snapshot of a pattern created by sequential update, where sites are updated in an ordered way: sites of the first line of the lattice are firstly updated from left to right, then sites of the second line, and so forth.
	This updating schema seems to conduce the system to a similar labyrinth phase found in the asynchronous case but with a lower density, $\rho_{\infty}=0.408$.
	This lower number of alive sites reflects the existence of domains of dead sites that prevents the establishment of denser structures of alive sites.
	Curiously, the value $\rho_{\infty}=0.408$ is very close to the site percolation threshold $p_c=0.407$ for a square lattice with Moore neighborhood~\cite{Percolation}.

	
	

\section{Mean-field analysis}

	
	Although we are studying bi-dimensional rules, it is interesting to consider if mean field analysis can give any insight for the $2D$ case.
	Each rule define an unique automaton and is given by $R[s,h]$.
	The rule $R[s,h]$ specifies the next state of a cell in the state $s$ ($s=0$ for a dead site, $s=1$ for an alive site) with $h$ alive neighbors.
	
	In this approach, we want to calculate the density of alive cells $\rho(t)$ \cite{Bagnoli1991,Malarz1998}.
	By neglecting spatial correlations, the time evolution expression for $\rho(t)$ is given by
	
	\begin{equation}
		\rho(t+1)=\sum_{s=0}^{1} \sum_{h=0}^{8} R[s,h]P_t(s,h),
	\label{MF}
	\end{equation}
	
	\noindent where the term $P_t(s,h)$ is the probability of a site $s$ having $h$ alive neighbors at time $t$.
	
	By using $P_t(s,h)=P(s)P(h,t)$, it can be shown that Eq.~(\ref{MF}) results in 
	
\begin{eqnarray}
 M(\rho) = \rho(t+1)  = (1-\rho(t)) & \sum_{h=0}^{8}R[0,h]P(h,t)&     \nonumber \\
											 + \rho(t) &\sum_{h=0}^{8}R[1,h]P(h,t).&
	\label{MAP}
\end{eqnarray}

 \noindent Here, $P(h,t)$ is the probability of finding a number $h$ of alive neighbors, which is assumed to follow a binomial distribution: $P(h,t)=C^8_h \rho(t)^h(1-\rho(t))^{8-h}$.

	The Life's rule is given by $R[0,3]=1$, $R[1,2]=1$, $R[1,3]=1$ and $R[s,h]=0$ for other values of $s,h$.
	The application of this rule in Eq.~(\ref{MAP}) produces the map:
	
	\begin{equation}
		M(\rho)=28\rho(t)^3(1-\rho(t))^5(3-\rho(t)).
	\label{MAP_LIFE}
	\end{equation}

	\noindent This map is a polynomial dominated by a order-3 term $\rho(t)^3$ when $\rho \rightarrow 0$ (which was confirmed numerically in \cite{Malarz1998}), and we call it a order-3 rule.
	This limit behavior is important to determine the stable points of the return map in Eq.~(\ref{MAP_LIFE}).
	
	 The map of Eq.~(\ref{MAP}) can be calculated for any rule and gives the temporal evolution of the density $\rho(t)$ given $\rho(0)$.
		It is valid and describes very well CA rules embedded in random networks with eight neighbours. 
		However, it fails in $2D$ for several important rules, like Life. This induced researchers to think that it is useless to understand complex rules. 
		We have shown~\cite{Reia_Kinouchi2014} that this is not the case: the mean-field fixed points can aid to define a control parameter $\sigma$ that locate complex rules inside the metastable region of a phase transition. 
		Complex rules are complex because in $2D$ there is nucleation, competition and coexistence of domains (formed by the dense fixed point phase) with the zero phase.

	\section{Results}
		
	In this context, we focus our attention only on the order-3 rules from Eq.~(\ref{MAP}) because they are the most dynamically interesting.
	These rules generally have two stable fixed points: $\rho=0$ and $\rho=\rho^*$.
	Reia and Kinouchi~\cite{Reia_Kinouchi2014} have claimed that $2D$ system presents alive and dead domains with similar densities as the mean-field fixed points.	
	Because of that, we study the subset of $6144$ (including Life) order-3 automata in the space of $262144$ outer-totalistic CA rules to analyze the coexistence and competition between these domains.  
	From this subset, $2643$ rules were excluded from our analysis by not having a stable fixed point $\rho^*$. Thus, our sample is composed of $3501$ rules.

\begin{figure}[b!]
	\centering
		\subfigure[]{\includegraphics[width=0.48\textwidth]{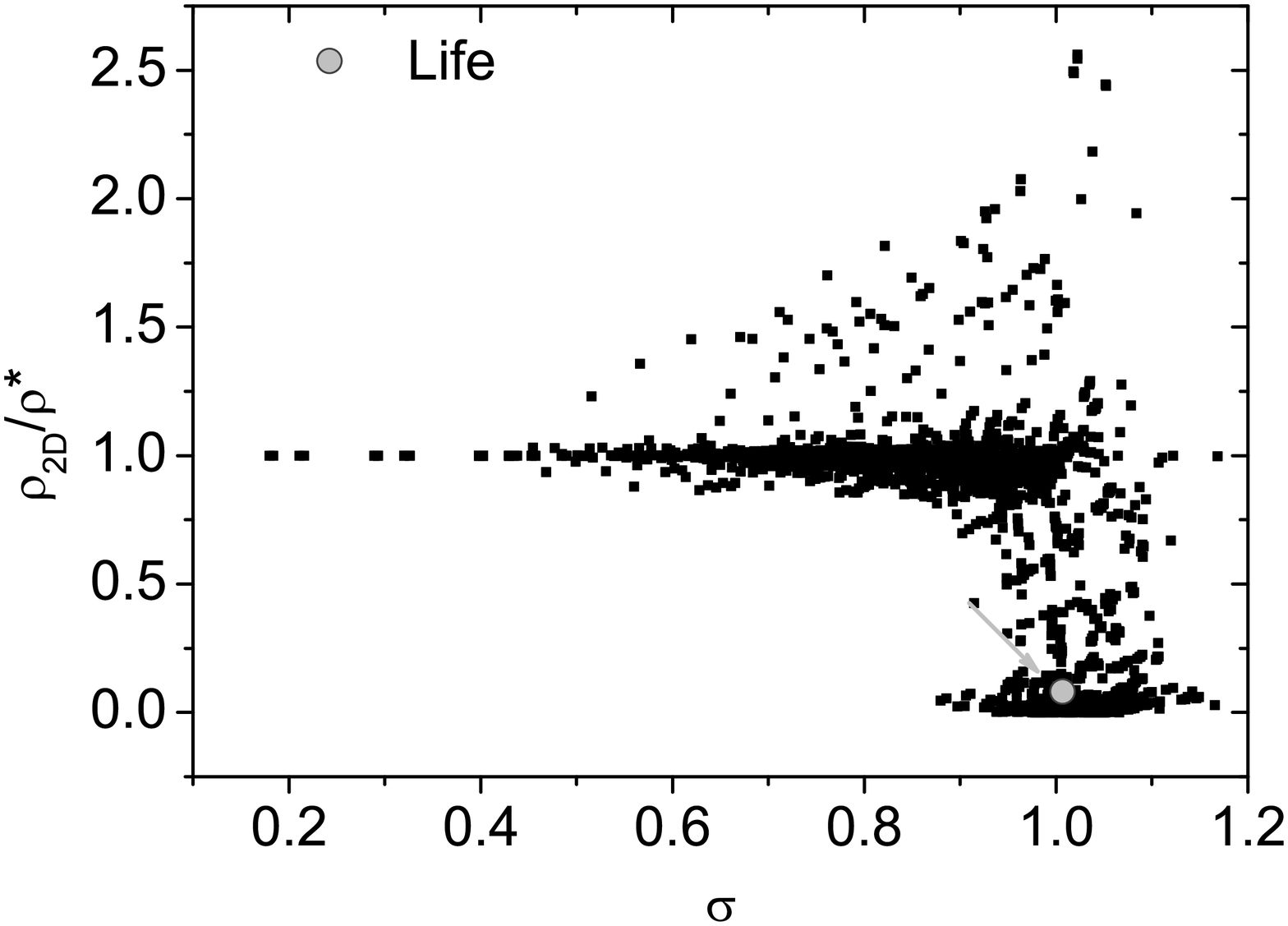}\label{synchronous_T}}
		\subfigure[]{\includegraphics[width=0.48\textwidth]{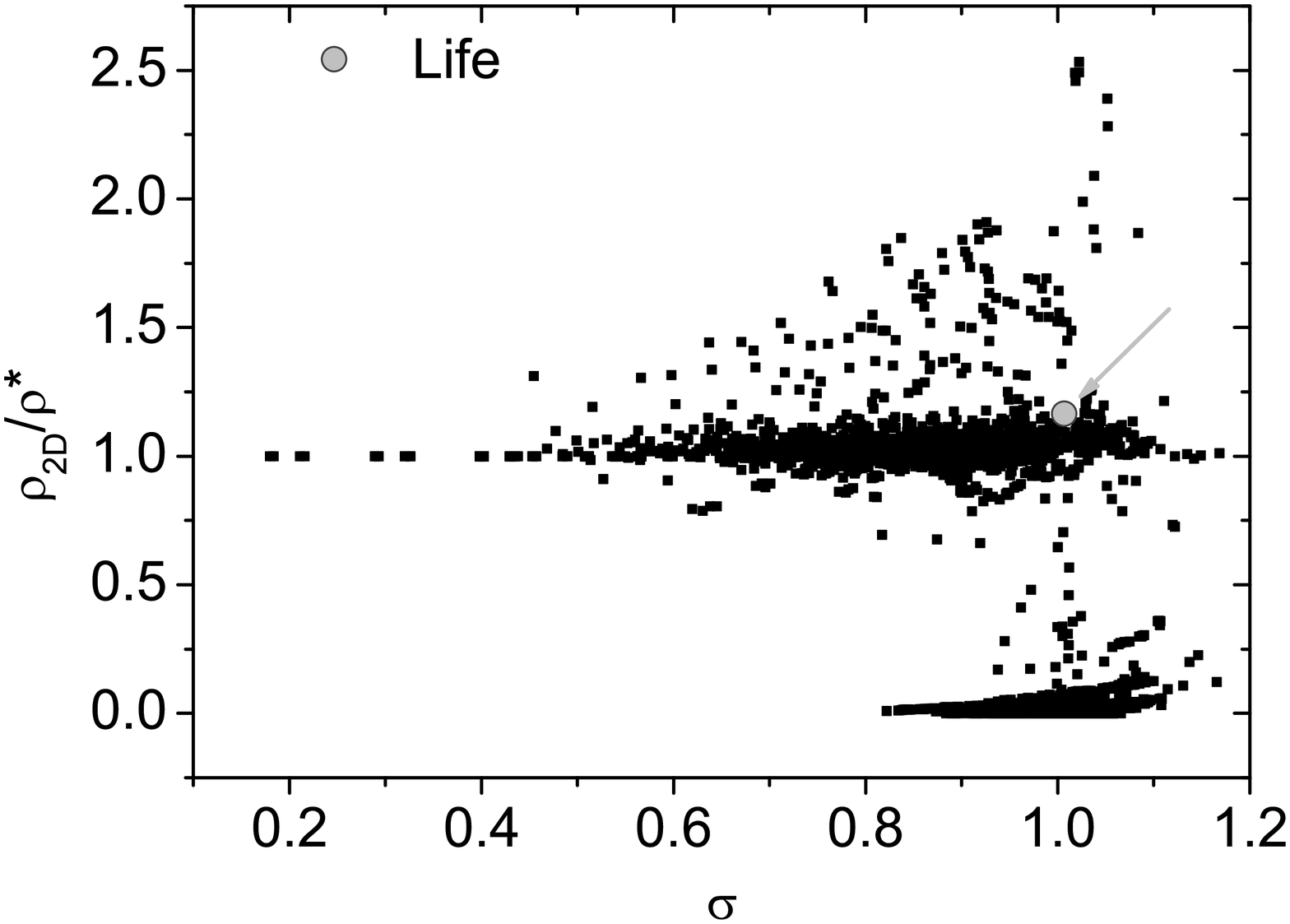}\label{asynchronous_T}}
		\subfigure[]{\includegraphics[width=0.48\textwidth]{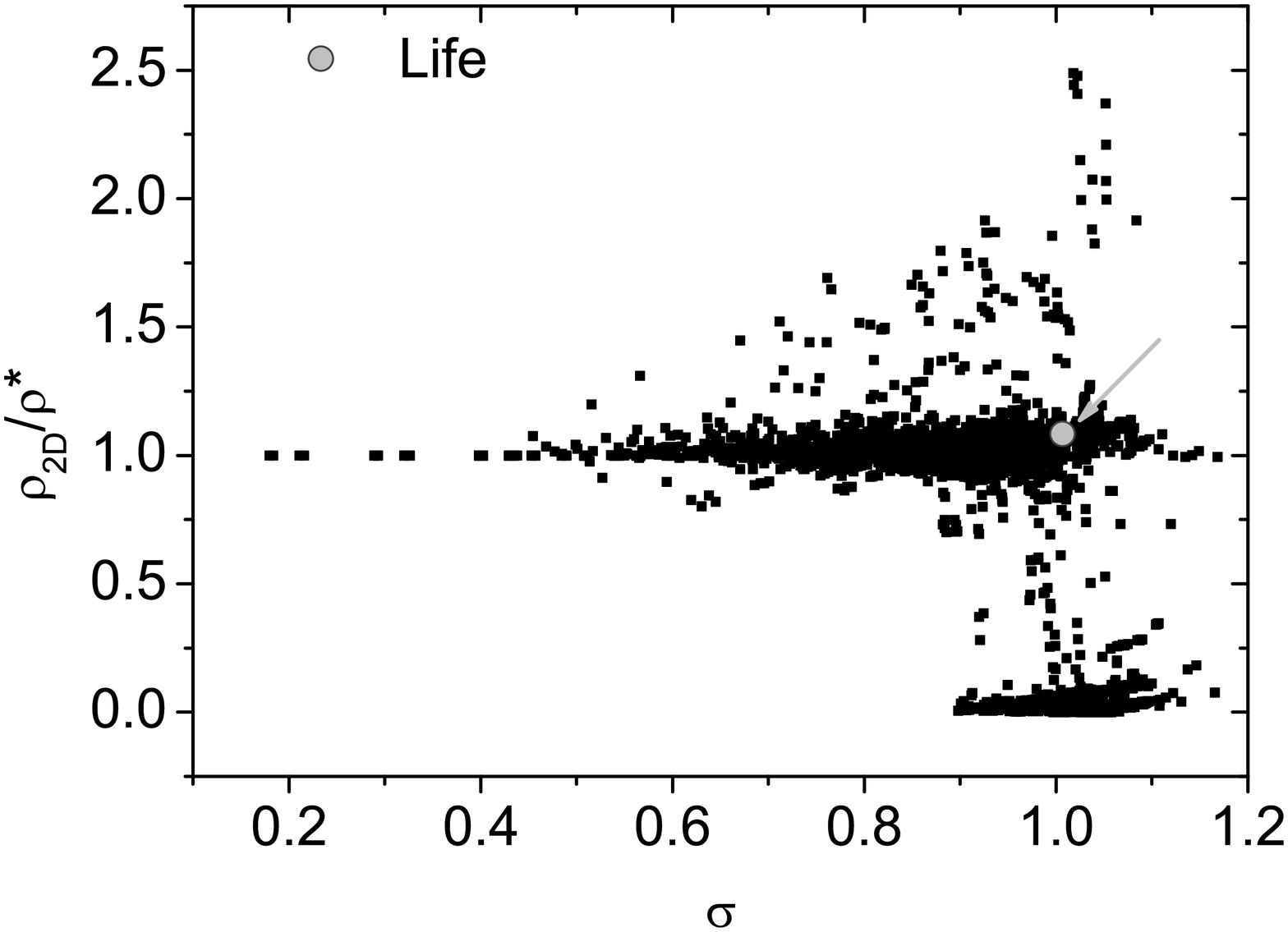}\label{sequential_T}}
		\caption{Relation between $\rho_{2D} / \rho^*$ and $\sigma$. Results were obtained for $L=100$ and averaged over $30$ runs. We chose $\rho(0)=\rho^*$ to see the effect of $\sigma$ on competition of alive and dead cells in $2D$ with synchronous update in panel~\ref{synchronous_T}, with asynchronous update in panel~\ref{asynchronous_T}, and sequential update in panel~\ref{sequential_T}. The Life $\sigma=1.006$ and its position is marked with a large grey dot in each panel (pointed by the grey arrow).}
	\label{fig4}
\end{figure}

	In order to see if the square lattice regards any resemblance with mean-field predictions, we plot $\rho_{2D}/\rho^*$ at the stationary state as function of a control parameter $\sigma$ in Fig.~\ref{fig4}, where $\rho_{2D}$ is density of alive sites in the lattice.
	The parameter $\sigma$ was devised to describe the growth rate of the vacuum phase at the interface of domains.
	Considering a square lattice composed of alive bubbles with density $\rho=\rho^*$ and dead bubbles with $\rho=0$, one can show that density at interface between both bubbles is $\rho^*/2$.
	Hence, the expansion and growing of dead sites can be inferred by defining the zero phase growth rate at interface~\cite{Reia_Kinouchi2014}:
	
	\begin{equation}
		\sigma=\frac{1-M(\rho^*/2)}{1-\rho^*/2}.
	\label{SIGMA}
	\end{equation}
	
	The zero phase expands for $\sigma>1$ and contracts for $\sigma<1$.
	The critical value is $\sigma_c=1$ and is at the transition region of Fig.~\ref{fig4}.
	Fig.~\ref{fig4} shows a large number of rules where $2D$ correlations are weak, which results in $\rho_{2D}=\rho^*$.
	For the other rules, we distinguish three cases: (I) spatial correlations promote overactivy, resulting in $\rho_{2D}/\rho^*>1$; (II) the mean-field prediction overestimates the density of alive cells, which results in a coexistence and competition between alive and dead domains with $0<\rho_{2D}/\rho_{MF}<1$; (III) despite the stable $\rho^*$ predicted by mean-field calculations, bubbles of zero phase expand over the entire lattice, resulting in an absorbing phase with $\rho_{2D}=0$.
		
	What we also see in Fig.~\ref{fig4} is that different updating schemas do not affect the phase transition as it does to individual $2D$ structural patterns.
	In fact, we observe that the destruction of structural patterns reduces the number of rules exhibiting coexistence and competition between domains (case(II)).
	As a consequence, although the transition region is less populated in asynchronous~\ref{asynchronous_T} and sequential~\ref{sequential_T} than synchronous~\ref{synchronous_T} update, the dead absorbing state and the high density state regimes can be identified by using $\sigma$.

	

\begin{figure}[b!]
	\centering
		\subfigure[]{\includegraphics[width=0.46\textwidth]{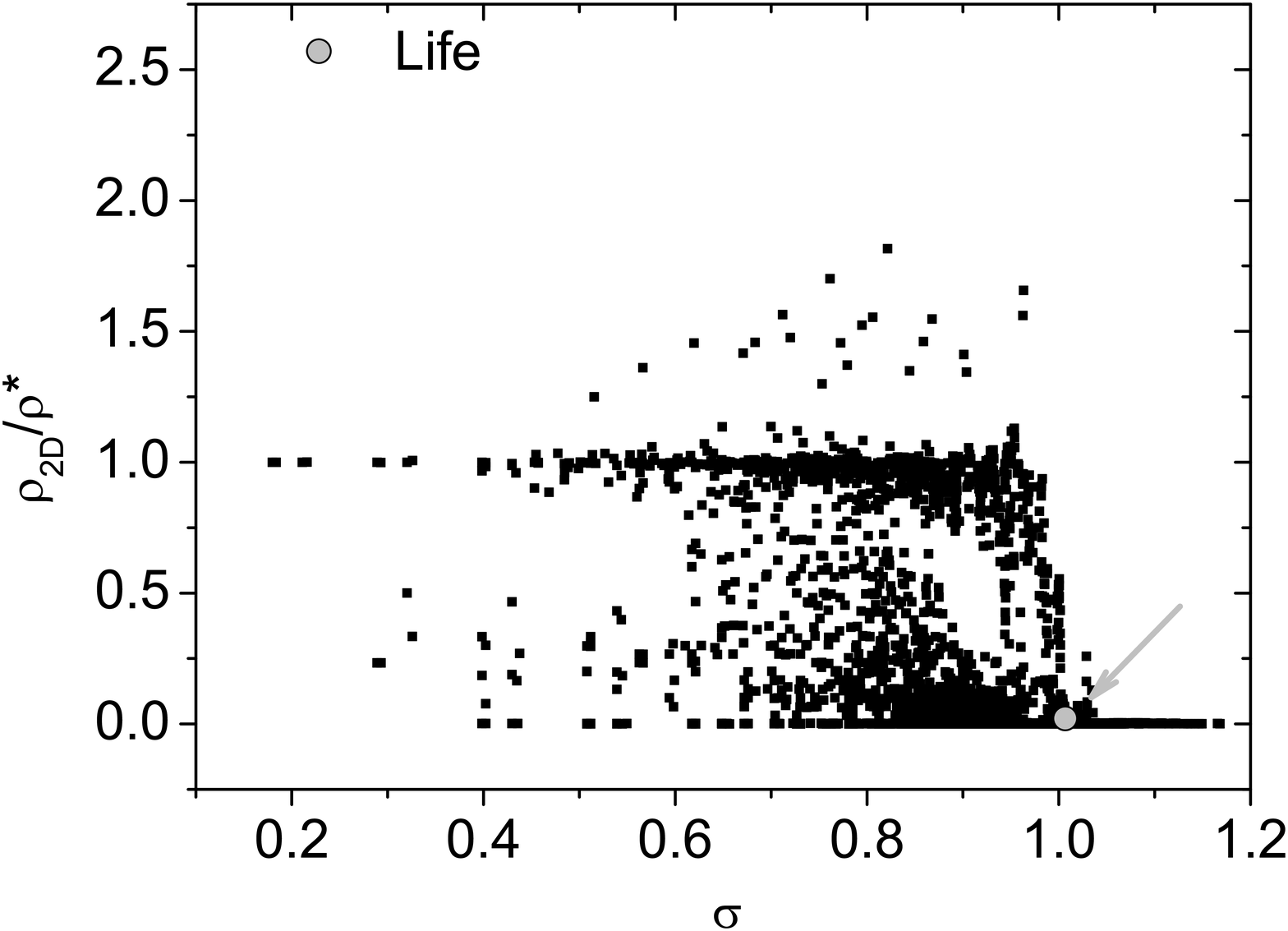}\label{synchronous_05}}
		\subfigure[]{\includegraphics[width=0.46\textwidth]{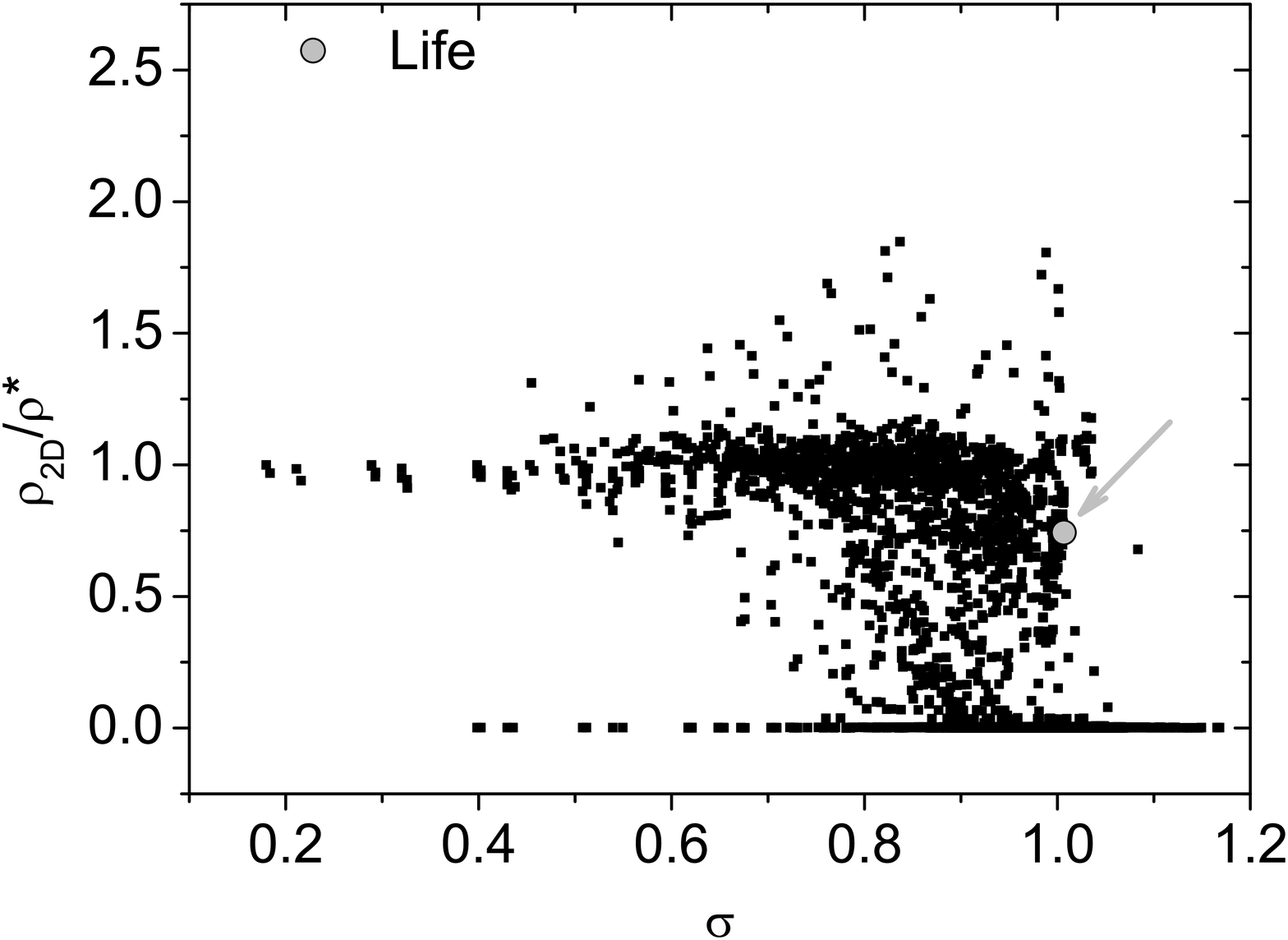}\label{asynchronous_05}}
		\subfigure[]{\includegraphics[width=0.46\textwidth]{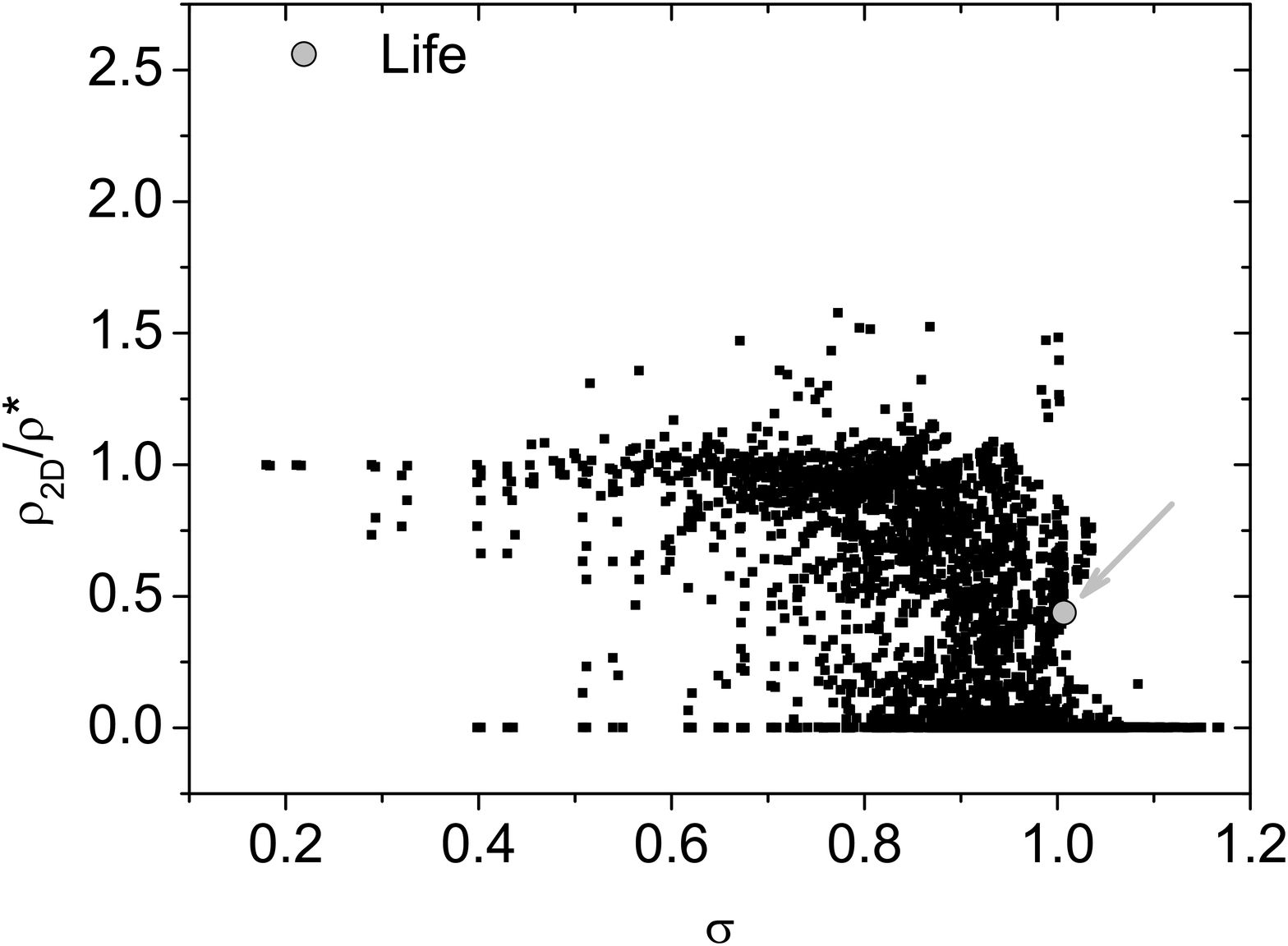}\label{sequential_05}}
		\caption{Relation between $\rho_{2D} / \rho^*$ and $\sigma$. Results were obtained for $L=100$ and averaged over $30$ runs. We chose $\rho(0)=0+ \Delta \rho$, where $\Delta \rho=0.05$, to see the effect of $\sigma$ on competition of alive and dead cells in $2D$ with synchronous update in panel~\ref{synchronous_05}, with asynchronous update in panel~\ref{asynchronous_05}, and sequential update in panel~\ref{sequential_05}. The Life $\sigma=1.006$ and its position is marked with a large grey dot in each panel (pointed by the grey arrow).}
	\label{fig05}
\end{figure}

	We have mentioned that the order-3 rules generally have two stable fixed points.
	The results shown in Fig.~\ref{fig4} report the phase transition found for $\rho(0)=\rho^*$.
	In Fig.~\ref{fig05}, we show the dependence of $\rho_{2D}/\rho^*$ on $\sigma$ when the initial density is $\rho(0)=0+\Delta \rho$, where $\Delta \rho=0.05$, under the three update schemas.
	We observe that the transition region is broadened and shifted to left in all schemas.
	We can note that the three cases already discussed are also seen here.
	However, we also observe that the low initial density reduces the number of rules having $\rho_{2D}/\rho^*>1$ (case (I)), mainly in synchronous (\ref{synchronous_05}) and sequential (\ref{sequential_05}) updates.
	This occurs because the high number of bubbles of zero phase found at the beginning of the simulation remains stable during the dynamics.
	Then, the growth of alive sites is constrained and there is a lower number or rules exhibiting overactivity.
	
	Considering that the low initial density acts in favor of bubbles of zero phase, one could expect that the dead sites would prevail over alive domains in a scenario of coexistence and competition.
	This expectation is confirmed by the higher number of rules having an absorbing phase $\rho_{2D}=0$ (case(III)) in Fig.~\ref{fig05}.
	In fact, it also explains the higher number of rules in region $0<\rho_{2D}/\rho^*<1$ (case(II)).
	Specifically, the initial density is too low that even when $\sigma<1$ the rules are attracted to the stable fixed point $\rho=0$.
	As a consequence, dead domains become stable and the value of $\rho_{2D}/\rho^*$ of these rules decrease.

\begin{figure}[t!]
	\centering
	\subfigure[]{\includegraphics[width=0.15\textwidth,frame]{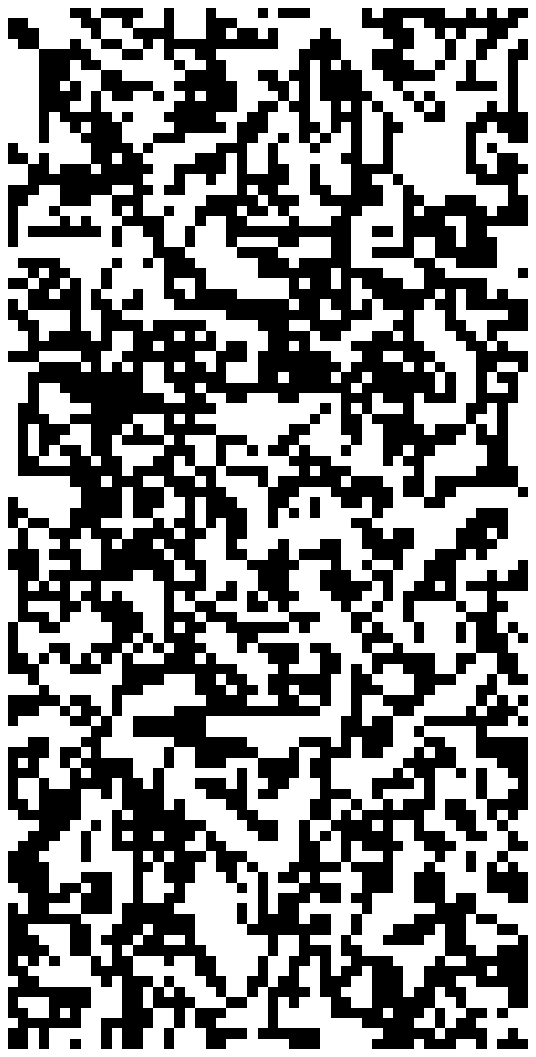}\label{B4678S2345_PARALELO}}
		\subfigure[]{\includegraphics[width=0.15\textwidth,frame]{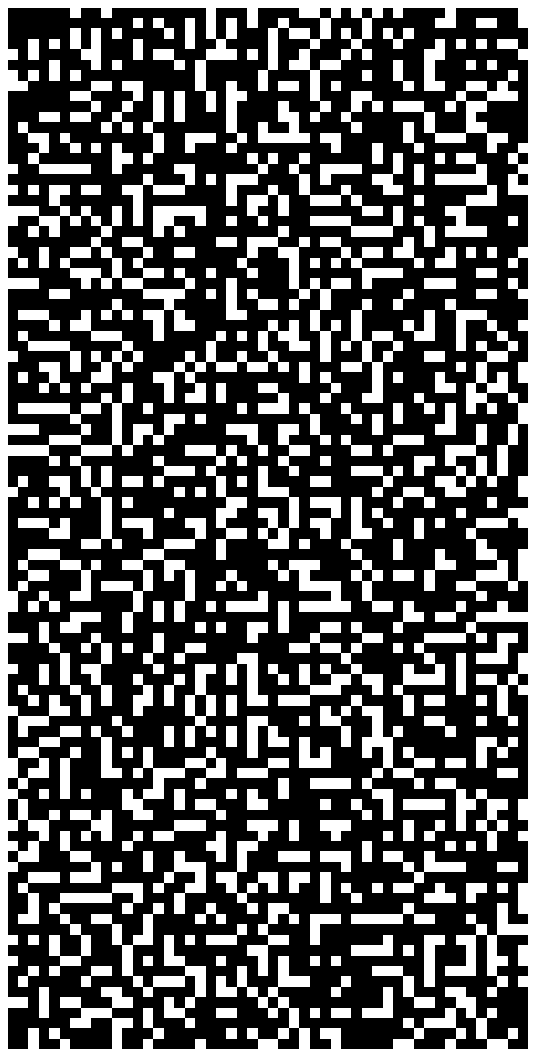}\label{B4678S2345_ASSINCRONO}}
		\subfigure[]{\includegraphics[width=0.15\textwidth,frame]{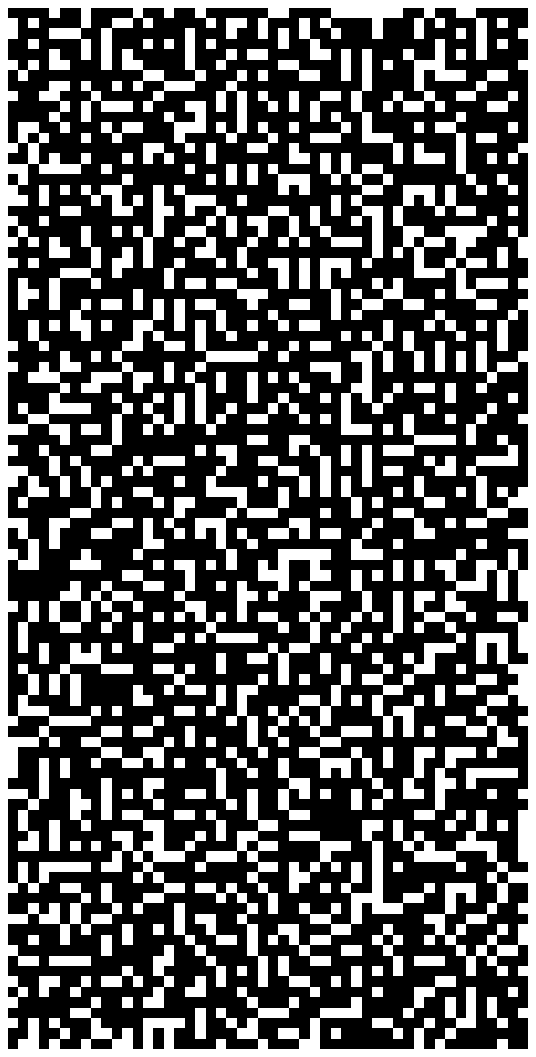}\label{B4678S2345_SEQUENCIAL}}
		\subfigure[]{\includegraphics[width=0.15\textwidth,frame]{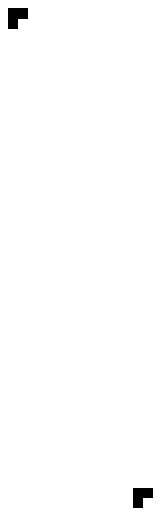}\label{B4678S2345_PARALELO_0}}
		\subfigure[]{\includegraphics[width=0.15\textwidth,frame]{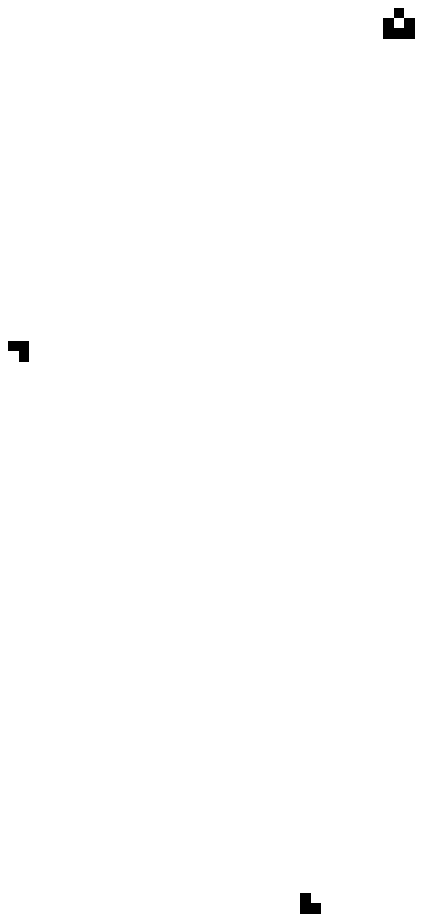}\label{B4678S2345_ASSINCRONO_0}}
		\subfigure[]{\includegraphics[width=0.15\textwidth,frame]{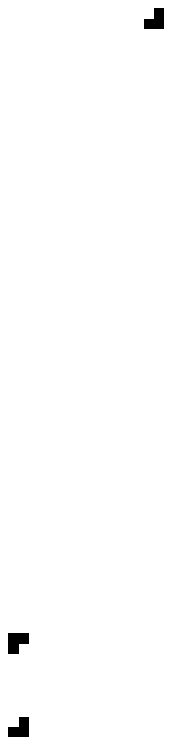}\label{B4678S2345_SEQUENCIAL_0}}
		\subfigure[]{\includegraphics[width=0.15\textwidth,frame]{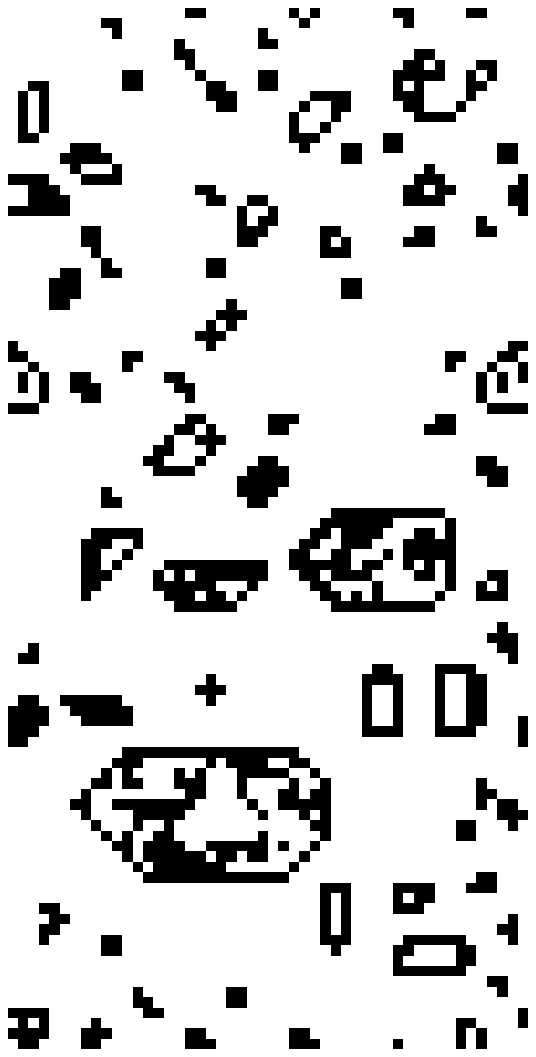}\label{B4678S2345_PARALELO_020}}
		\subfigure[]{\includegraphics[width=0.15\textwidth,frame]{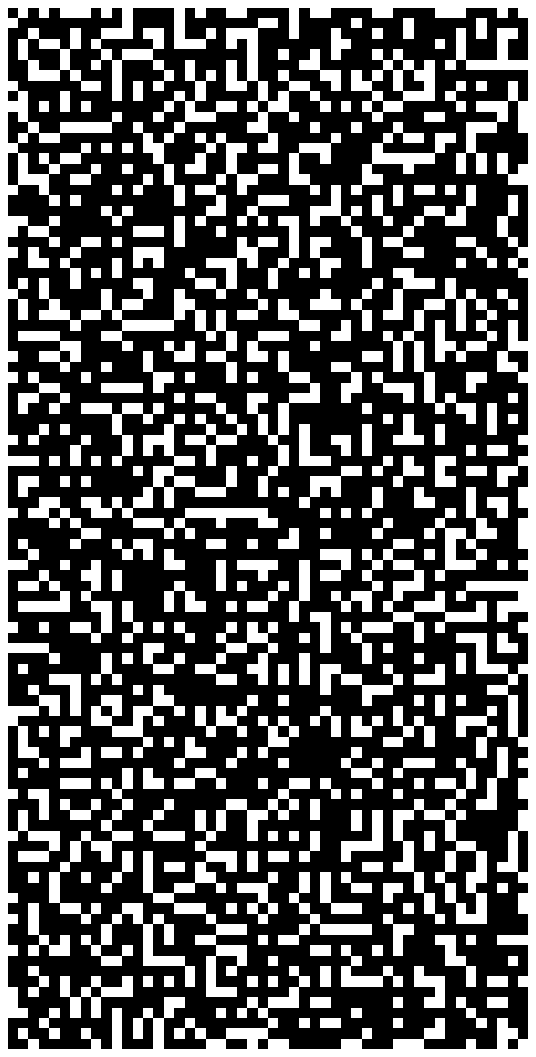}\label{B4678S2345_ASSINCRONO_020}}
		\subfigure[]{\includegraphics[width=0.15\textwidth,frame]{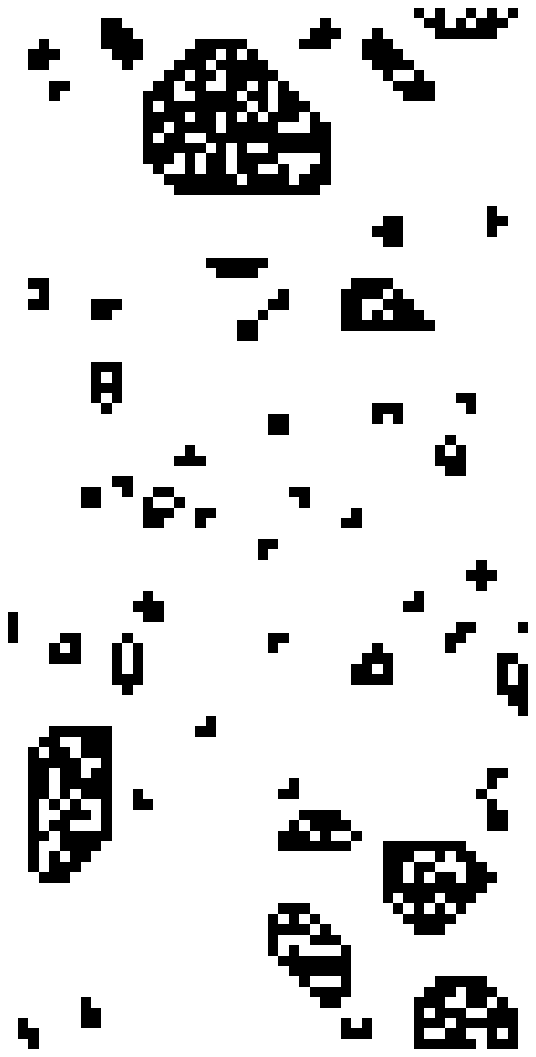}\label{B4678S2345_SEQUENCIAL_020}}
		\caption{Snapshot at steady state of rule B4678S2345 with synchronous~\ref{B4678S2345_PARALELO}, asynchronous~\ref{B4678S2345_ASSINCRONO} and sequential~\ref{B4678S2345_SEQUENCIAL} updates in a $2D$ lattice with $L_m \times L_n =100 \times 50$. The system evolved to steady state from a randomly initial condition with $\rho(0)=\rho^*$. Figs.~\ref{B4678S2345_PARALELO_0}, \ref{B4678S2345_ASSINCRONO_0} and \ref{B4678S2345_SEQUENCIAL_0} show steady states for the three schemas when $\rho(0)=0 + \Delta \rho$, where $\Delta \rho = 0.05$. Figs.~\ref{B4678S2345_PARALELO_020}, \ref{B4678S2345_ASSINCRONO_020} and \ref{B4678S2345_SEQUENCIAL_020} show steady states for updates synchronous, asynchronous and sequential, respectively, when $\rho(0)=0.200$.  }
	\label{B4678S2345_LOW_PATTERN}
\end{figure}

	As a matter of illustration, we show in Fig.~\ref{B4678S2345_LOW_PATTERN} $2D$ steady states for a rule with $\sigma<1$ whose value $\rho_{2D}/\rho^* \approx 1$ decay when the initial density is decreased.
	This rule is defined by $B4678S2345$, which has $\sigma=0.956$ and two stable fixed points: $\rho=0$ and $\rho^*=0.615$.
	We see in Figs.~\ref{B4678S2345_PARALELO}, \ref{B4678S2345_ASSINCRONO} and \ref{B4678S2345_SEQUENCIAL}, the steady state with synchronous, asynchronous and sequential updates respectively.
	Note that the initial condition $\rho(0)=\rho^*$ allows a strong coexistence and competition between alive and dead domains in a synchronous updating, resulting in $\rho_{\infty}=0.491$. 
	As already discussed for Life, the asynchronous and sequential updates acts in favor of alive sites.
	Consequently, alive sites prevail over dead domains and results in a stationary density $\rho_{\infty}=0.678$ in both cases.

	We see in Figs.~\ref{B4678S2345_PARALELO_0}, \ref{B4678S2345_ASSINCRONO_0} and \ref{B4678S2345_SEQUENCIAL_0} the steady states for the three updates when $\rho(0)=0+\Delta \rho$, where $\Delta \rho=0.05$.
	We can note that the high number of dead domains found at the beginning of the dynamics expands over the entire lattice.
	This results in a small number of nuclei of alive sites, which results in $\rho_\infty \approx 0$ in the three schemas.
	This rule illustrates the underlying properties of the $\sigma<1$ rules that broaden and shift the transition region to left when the stability of the absorbing steady state is studied in Fig.~\ref{fig05}.

	The steady states of the system when an intermediary value $\rho(0)=0.200$ is assigned to the initial density is seen in Figs.~\ref{B4678S2345_PARALELO_020}, \ref{B4678S2345_ASSINCRONO_020} and \ref{B4678S2345_SEQUENCIAL_020}, under the three update schemas respectively.
	These snapshots depicts not only the strong coexistence and competition between dead and alive domains, which can be disturbed just by increasing or decreasing the initial density, but also the effect of the update in this subtle equilibrium.
	In Fig.~\ref{B4678S2345_PARALELO_020} we observe the formation of stable structures like ``crystals'' (or ``diamonds'') of alive sites that persist until the end of the synchronous dynamics, with $\rho_\infty=0.170$.
	In Fig.~\ref{B4678S2345_ASSINCRONO_020} the asynchronous update destroyed these typical structures and favored the spread of alive domains, resulting in $\rho_\infty = 0.683$.
	Curiously, Fig.\ref{B4678S2345_SEQUENCIAL_020} shows that sequential update promotes the establishment of stable structures of alive sites.
	However, in contrast to synchronous case, the lattice is poorly populated by alive domains, allowing the presence of greater regions of dead sites, but these nuclei of alive sites are denser than in synchronous case, which results in $\rho _\infty = 0.140$.

	
	
	
	

	It is noteworthy the results from Fates~\cite{Fates2010}.
	In this work, the author studies the dependence of the steady state density $\rho_{\infty}$ on the initial density $\rho(0)$ and on the degree of desynchronisation of the update.
	He finds that $\rho_{\infty}$ of Life is not related to $\rho(0)$ when the dynamical update is asynchronous.
	Our results suggest this is not the case for all automata rules.
	The snapshots at the steady state of the rule B4678S2345 showed in Fig.~\ref{B4678S2345_ASSINCRONO} and Fig.~\ref{B4678S2345_ASSINCRONO_0} reveals that the initial density affects the final pattern in the lattice and the final density.
	This may happen to the asynchronous rules that present a decay from $\rho_{2D}/\rho^* \approx 1$ in phase transition (Fig.~\ref{asynchronous_T}) when the initial density is changed (Fig.~\ref{asynchronous_05}), as already discussed.
	Despite that, the general aspects of phase transition remains qualitatively the same.

\section{Conclusion}

	We study the effects of asynchronous and sequential updates in the dynamics of 3501 CA in a square lattice.
	By using Life as example, we see that updates changes destroys characteristic structures of this automaton and are favorable to the spread of alive sites.
	We have shown that the first-order phase-transition found with the use of the recently proposed control parameter $\sigma$ is observed not only in the synchronous but also in asynchronous and sequential updates.
	
	The parameter $\sigma$ measures the growth rate of the vacuum phase at the domains interface.
	Both updates schema also present a transition from a high density state to a dead absorbing state around $\sigma_c=1$, which indicates that the control parameter $\sigma$, although heuristic, seems to be robust.
	Indeed, this transition was found to be qualitatively the same when the initial density was changed.
	The fact that the transition is robust and that complex rules lie near $\sigma_c=1$ illustrates the idea that, in the multiverse of CA, complex universes lie inside a phase transition region.


\section{Acknowledgments}
	The authors thank Bernardo A. Huberman for discussions.
	S. Reia thanks CAPES for the financial support and Ariadne A. Costa for useful conversations.
	O. Kinouchi acknowledges support from CNPq and CNAIPS-USP.

\bibliographystyle{apsrev}
\bibliography{references}

\end{document}